\documentclass[prl,aps,amsmath,twocolumn,amssymb,superscriptaddress]{revtex4}

\usepackage{graphicx}

\def\str{\mathop{\rm str}}

\def\erf{\mathop{\rm erf}}

\renewcommand{\Re}{\mathop{\rm Re}}

\begin{document}

\title{Anderson localization of a Majorana fermion}

\author{D.~A.~Ivanov}
\affiliation{Institute for Theoretical Physics, ETH Z\"urich, 8093 Z\"urich, Switzerland}
\affiliation{Institute for Theoretical Physics, University of Z\"urich, 8057 Z\"urich, Switzerland}

\author{P.~M.~Ostrovsky}
\affiliation{Max Planck Institute for Solid State Research, Heisenbergstr.\ 1,
70569 Stuttgart, Germany}
\affiliation{L.~D.~Landau Institute for Theoretical Physics, 142432 Chernogolovka, Russia}

\author{M.~A.~Skvortsov}
\affiliation{L.~D.~Landau Institute for Theoretical Physics, 142432 Chernogolovka, Russia}
\affiliation{Moscow Institute of Physics and Technology, 141700 Moscow, Russia}

\date{June 29, 2013}

\begin{abstract}
Isolated Majorana fermion states can be produced at the boundary of a topological
superconductor in a quasi-one-dimensional geometry. If such a superconductor is
connected to a disordered quantum wire, the Majorana fermion is spread into the wire,
subject to Anderson localization. We study this effect in the limit of a thick wire
with broken time-reversal and spin-rotational symmetries. With the use of a
supersymmetric nonlinear sigma model, we calculate the average local density of states
in the wire as a function of energy and of the distance from the interface with
the superconductor. Our results may be qualitatively explained by the repulsion
of states from the Majorana level and by Mott hybridization of localized states.
\end{abstract}

%\pacs{
%74.45.+c, %Proximity effects; Andreev reflection; SN and SNS junctions}
%73.20.Fz, %Weak or Anderson localization
%73.21.Hb  %Electron states and collective excitations in: Quantum wires
%}

\maketitle

\paragraph{1. Introduction.---}
In recent years, search for Majorana electron levels in solid-state systems has
intensified, motivated by their potential use in quantum computing \cite{beenakker:1.11}.
Several experimental groups reported indications of Majorana fermions in
hybrid superconductor-semiconductor systems \cite{mourik:12,das:12,deng:12,rokhinson:12},
which further stimulated theoretical studies of Majorana fermions in superconducting
proximity structures.
One of the interesting theoretical questions relevant for the analysis of
the recent experiments is the role of disorder in the structure of the
electronic spectrum of systems with Majorana fermions. This topic was subject of
several recent papers, which addressed transport properties of hybrid
systems with Majorana states \cite{akhmerov:11,wimmer:11,liu:12,pikulin:12}.

An alternative approach to disordered systems is in terms of correlation
functions, the simplest of which is the average local density of states (LDOS).
If the disorder is strong enough, the electronic states are localized \cite{anderson:58},
including the Majorana fermion. At the same time, the LDOS
at low energies is modified, due to the level repulsion from the
Majorana state (this effect was studied in detail in the zero-dimensional
random-matrix limit \cite{bocquet:00,ivanov:02}).
In the case of a thick quasi-one-dimensional wire
with a large number of channels, the modification of the density of states
in the presence of a localized Majorana level can be accessed with the
use of a supersymmetric nonlinear sigma model \cite{efetov:1-83}. This approach
was developed in Ref.~\onlinecite{skvortsov:12} for the case of a non-topological
superconducting system without Majorana fermion.

In the present paper, we adapt the calculation of Ref.~\onlinecite{skvortsov:12}
to the case of a hybrid system consisting of a topological superconductor
with a Majorana level and a wire with both time-reversal and spin symmetries broken.
As a result, we find the spatial structure of the localized Majorana state
and the modification of the density of states at low energies due to the level
repulsion from the Majorana level. The localization of the Majorana level
is similar to the localization of states in the bulk of the wire,
and its level-repulsion effect extends to the Mott length scale. We
further explain our results in terms of the Mott
hybridization of localized states \cite{mott:68:70,ivanov:12}
and compare our findings to those in Ref.~\onlinecite{skvortsov:12}.

\begin{figure}
\includegraphics[width=.35\textwidth]{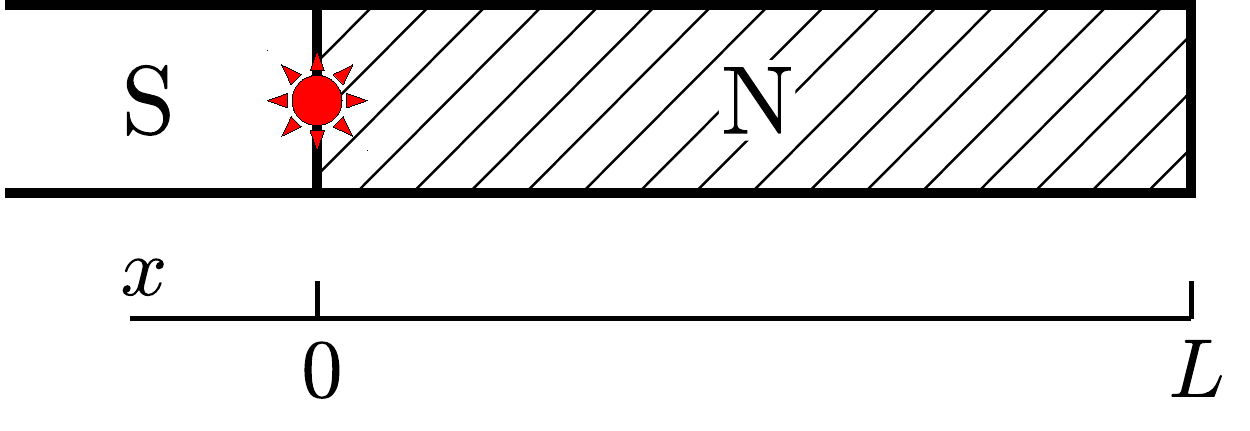}
\caption{(Color online)
A disordered quantum wire (N) of length $L$ coupled to a topological
superconductor (S) at $x=0$. The star marks the location of a Majorana fermion
state. This state spreads at the localization length scale $\xi$ into
the wire and at the superconducting coherence length (which is assumed
to be much shorter than $\xi$) into the superconductor.}
\label{fig:setup}
\end{figure}

\paragraph{2. Model.---}
The model we consider is identical to that described in Ref.~\onlinecite{skvortsov:12},
with the only difference that the superconductor is now
of a topological type hosting a Majorana fermion
at the interface with the normal metal \cite{beenakker:2.11}.
The setup is schematically shown in Fig.~\ref{fig:setup}.
The wire has length $L$ and the number of channels $N \gg 1$.
Both the spin-rotational and time-reversal
symmetry are assumed to be broken in the wire
(at the length scales shorter than all other length
scales of the theory). It therefore belongs to the symmetry class A
(in Cartan notation \cite{zirnbauer:96})
and to the model IIb in Efetov's classifcation \cite{efetov:97}. The superconductor
is also assumed to have both time-reversal and spin-rotational symmetry
broken and to be in a topological phase \cite{beenakker:2.11}.
This superconducting symmetry class is usually
denoted as B, D or BD \cite{zirnbauer:96,ivanov:02,bocquet:00,altland:97}.
We prefer to call this symmetry class B to emphasize the presence of
a Majorana state at the boundary.
An example of a microscopic Hamiltonian leading to such a symmetry can
be found in Refs.~\onlinecite{lutchyn:10,oreg:10}.

We assume quantum coherence in the wire, which leads to Anderson localization
at the localization length $\xi = 2\pi \nu A D$
(here D is the diffusion coefficient in the wire, $\nu$ is the density of states,
and $A$ is the wire cross section). Localization in the wire defines
the energy scale $\Delta_\xi = D/\xi^2$ (of the order of the level spacing and
of the Thouless energy at the length $\xi$).
In short wires ($L\ll \xi$), another energy scale comes
into play: the level spacing in the wire,  $\delta=(2\nu A L)^{-1}$
\cite{com-delta}. The localization of the
Majorana fermion and the related level-repulsion effects occur at distances
$x \lesssim \xi$ from the interface [up to logarithmic corrections, see the
discussion of the Mott length (\ref{Mott-length}) below] and at energies
$E \lesssim \max(\Delta_\xi,\delta)$ around the Fermi level,
as in the non-topological  case \cite{skvortsov:12}. We further assume
these conditions on $x$ and $E$.

The interface between such a superconductor and a wire may be described
in terms of a scattering matrix $2N{\times}2N$, which involves both
normal and Andreev reflections. We assume a large total Andreev conductance.
Additionally, we assume that the gap in the superconducting part
of the junction is much larger than the energy scales $\Delta_\xi$ and $\delta$.
Then, within the window of energies considered, we may neglect the energy
dependence of the Andreev scattering matrix (the limit of a broad ``Majorana
resonance'' \cite{law:09,ioselevich:12}). The symmetry properties of such
a scattering matrix are discussed in detail
in Refs.~\onlinecite{akhmerov:11,beenakker:2.11,pikulin:12}.
Similarly to the results of Ref.~\onlinecite{skvortsov:12}, we
find that the role of the Andreev interface, in this high-transparency
limit, is to fully suppress the soft localization modes (diffusons and cooperons)
incompatible with the symmetry of the superconductor. At such an interface,
the boundary conditions for the supersymmetric
nonlinear sigma model describing the wire take a particularly simple
form (see the following section).

\paragraph{3. Density of states: general formalism.---}
Our main object of study is the disorder-averaged local density of states
$\langle \rho_E(x) \rangle$ (normalized to the bulk value in the wire).
Following the procedure described  in Ref.~\onlinecite{skvortsov:12},
we express it as an expectation value in a
supersymmetric nonlinear sigma model \cite{efetov:97},
\begin{equation}
\langle \rho_E(x) \rangle =
\frac{1}{4} \Re \int \str \left( k\Lambda Q(x) \right) e^{-S[Q]} DQ \, .
\label{functional-integral}
\end{equation}
Here $Q$ is the supersymmetric $4{\times}4$ matrix acting in the product
of the Fermi-Bose (FB) and retarded-advanced (RA) two-dimensional spaces
and subject to the constraint $Q^2=1$ (this space of $Q$ matrices
corresponds to the symmetry class A \cite{zirnbauer:96}).
The matrix $\Lambda=\sigma_z^\text{RA}$ is the metallic saddle point
of the sigma model, and the supersymmetry-breaking matrix $k$
is defined as $k=\sigma_z^\text{FB}$ (we follow the notations
of Ref.~\cite{efetov:97}).
The supersymmetric action has the form
\begin{equation}
S[Q]=\frac{\pi \nu A}{4} \int dx\, \str \left[ D (\nabla Q)^2 + 4 i E \Lambda Q \right]\, .
\end{equation}
The boundary conditions on the matrix $Q$ are free at the end point $x=L$.
At the interface with the superconductor, under the assumptions
specified in the previous section, the matrix $Q$ is restricted to the symmetry
class of the combination of the symmetries in the wire and in the superconductor
(class B in our case).

As usual, we parameterize the $Q$ matrix by the eigenvalues of its RR
block, which are conventionally denoted $\lambda_F$ and $\lambda_B$ (for the
fermionic and bosonic eigenvalues, respectively) \cite{efetov:97}. The fermionic
parameter $\lambda_F$ takes values between $-1$ and $1$ (compact), and
the bosonic component $\lambda_B$ between $1$ and $+\infty$ (noncompact).
At the interface with a superconductor, the $Q$ matrix is reduced to that
of the symmetry class B or D (depending on the presence or absence
of a Majorana level), which constrains the FF block to two points only:
$\lambda_F=\pm 1$ \cite{bocquet:00,ivanov:02,skvortsov:12}. As derived
in Ref.~\onlinecite{skvortsov:12}, in the case of class D (no Majorana level),
the two components of the functional integral (\ref{functional-integral})
corresponding to $\lambda_F(x{=}0)=1$ and $\lambda_F(x{=}0)=-1$ must
be added with opposite signs. In Ref.~\onlinecite{ivanov:02}, it was remarked that
the difference between the presence and absence of the Majorana mode corresponds
to reversing the relative sign between the functional integrals over
the two disconnected components of the space of $Q$ matrices. Therefore,
we conclude that, in the case of class B (with the Majorana mode),
the average density of states is given by reversing the relative sign in Eq.~(28) of
Ref.~\onlinecite{skvortsov:12}:
\begin{equation}
  \langle\rho_E(x)\rangle
  =
  1 + \Re \int_1^\infty d\lambda_B \,
  \frac{\Psi(1,\lambda_B;x)  + \Psi(-1,\lambda_B;x)}{2}\,  ,
\label{rho-integral}
\end{equation}
with the wave function $\Psi(\lambda_F,\lambda_B; x)$
defined as \cite{efetov:2-83,ivanov:08,ivanov:09}
\begin{multline}
  \Psi(\lambda_F,\lambda_B;x) \\
  = e^{-2\tilde H x / \xi } \, (\lambda_B-\lambda_F) \, e^{-2\tilde H (L-x)/\xi}
  \, (\lambda_B-\lambda_F)^{-1} \, .
\label{full-wave-function-definition}
\end{multline}
The Hamiltonian $\tilde H$ governing the evolution of the wave function
is given by
\begin{equation}
  \tilde H  = \tilde H_B + \tilde H_F\, ,
  \label{tilde-H}
\end{equation}
where
\begin{subequations}
\begin{align}
\label{tilde-HB}
  & \tilde H_B
  = - \frac12 \frac{\partial}{\partial\lambda_B} (\lambda_B^2-1)
\frac{\partial}{\partial\lambda_B}
   + \frac{\kappa^2}{16} \lambda_B
  ,
\\
\label{tilde-HF}
  & \tilde H_F
  = - \frac12 \frac{\partial}{\partial\lambda_F} (1-\lambda_F^2)
\frac{\partial}{\partial\lambda_F}
  - \frac{\kappa^2}{16} \lambda_F
  ,
\end{align}
\end{subequations}
and we define $\kappa^2 = - 8iE/ \Delta_\xi$.

\paragraph{4. Short-wire limit.---}
In the limit of a short wire, $L\ll \min(\xi, \sqrt{D/E})$,
the gradient terms in the Hamiltonian (\ref{tilde-H})
may be neglected \cite{skvortsov:12}, and a simple
calculation leads to the well-known result for the
random-matrix ensemble of class B \cite{mehta:91,ivanov:02}:
\begin{equation}
\langle \rho_E(x) \rangle = 1 - \frac{\sin(2\pi E / \delta)}{(2\pi E / \delta)}
+ \delta \left( E / \delta \right)\, .
\end{equation}
The Majorana level reveals itself as a delta-function term in the
density of states. The integral weight of this delta peak equals
$1/2$, which reflects the Majorana nature of this electronic level.

\paragraph{5. Infinite-wire limit.---}
In the limit of an infinite wire, at low energies ($E \ll \Delta_\xi$),
the calculation can be performed
using the technique developed in Ref.~\onlinecite{ivanov:09}. Taking the limit
$L\to\infty$ in Eq.~(\ref{full-wave-function-definition}) results in
\begin{equation}
  \Psi(\lambda_F,\lambda_B;x)_{L\to\infty} = e^{-2(\tilde H_F + \tilde H_B)x/\xi} \, \Psi_0(\lambda_F,\lambda_B)\, ,
\label{psi-infinite}
\end{equation}
where
\begin{equation}
\Psi_0(\lambda_F,\lambda_B) = I_0(q)\, p K_1(p) + q I_1(q) \, K_0(p)
\label{zero-mode}
\end{equation}
(with $p=\kappa \sqrt{(\lambda_B+1)/2}$ and $q=\kappa \sqrt{(\lambda_F+1)/2}$)
is the zero mode of the one-dimensional sigma model in the unitary symmetry class
found in Ref.~\onlinecite{skvortsov:07}.

\begin{figure}
\includegraphics[width=.45\textwidth]{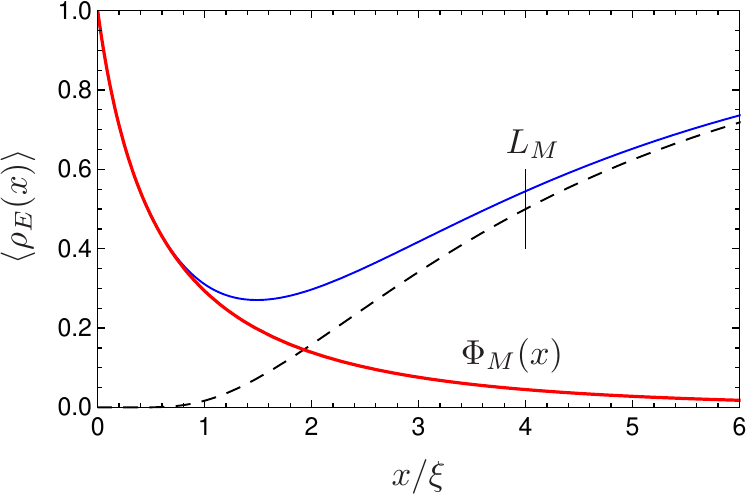}
\caption{
(Color online)
Sketch of the average LDOS $\langle \rho_E(x) \rangle$ 
in the infinite-wire limit ($L\to \infty$)
at a small finite energy $E$ as a function of $x$ (thin blue curve),
as given by Eq.~(\ref{LDOS-infinite}).
Close to the interface, it follows the profile of the Majorana mode
$\Phi_M(x)$ (thick red curve). At $x\sim L_M$,
the LDOS exhibits a crossover given by the erf function (dashed black line).}
\label{fig:rho}
\end{figure}

Since the bosonic and fermionic degrees of freedom separate, the calculation may
be further performed perturbatively in the fermionic sector (using the assumption
of $E\ll\Delta_\xi$) and by expanding in the basis of eigenstates of $\tilde H_B$ in
the bosonic sector \cite{ivanov:09}. The details of the calculation can be found
in the Supplemental Material \cite{supplemental}. The result has the form
\begin{multline}
  \langle \rho_E(x) \rangle
  =
  \Phi_M(x)
  +
  \frac{1}{2} \left[ 1 + \erf \left(\frac{x-L_M}{2 \sqrt{x\, \xi}} \right) \right]
\\
{}
  + \Phi_M(x) \, \pi \, \delta(E/\Delta_\xi)
\, .
\label{LDOS-infinite}
\end{multline}
The last $\delta$ term in Eq.~(\ref{LDOS-infinite}) corresponds to the localized
Majorana mode whose profile is given by
\begin{equation}
\Phi_M(x) = 2\pi \int_0^\infty k \, dk\, \frac{\sinh \pi k}{\cosh^2 \pi k}
(k^2 + 1/4) e^{-(x/\xi)(k^2+1/4)} ,
\label{Majorana-infinite}
\end{equation}
and
\begin{equation}
  L_M = 2\xi \ln (\Delta_\xi/E)
  \label{Mott-length}
\end{equation}
is the Mott length scale \cite{mott:68:70}. Note that the profile of the Majorana level
$\Phi_M(x)$ contributes not only to the delta peak at zero energy, but also
to the background LDOS at small finite energies (see our discussion of Mott
hybridization below). The function $\langle \rho_E(x) \rangle$ is plotted
in Fig.~\ref{fig:rho}.

As in the short-wire limit, one can verify that the integral weight of the
Majorana delta peak in the LDOS (\ref{LDOS-infinite}) equals $1/2$. The
{\em average}\ intensity of the Majorana mode decays away from the interface
at the length scale $4\xi$, similar to the statistics of a single localized
wave function \cite{gogolin:76,efetov:2-83,ivanov:09}.
On general grounds, we believe that the statistics of
the tails of this localized state is log-normal \cite{ivanov:12}
with the {\em typical}\ Majorana state decaying at the length scale $\xi$.
Note however that the exact
form of the average intensity of the Majorana state (\ref{Majorana-infinite})
differs from the two-point correlation function of the intensity of
a single localized state [Eq.~(75) of Ref.~\onlinecite{ivanov:09},
see also Refs.~\onlinecite{gogolin:76,efetov:2-83}].

\begin{figure}
\includegraphics[width=.3\textwidth]{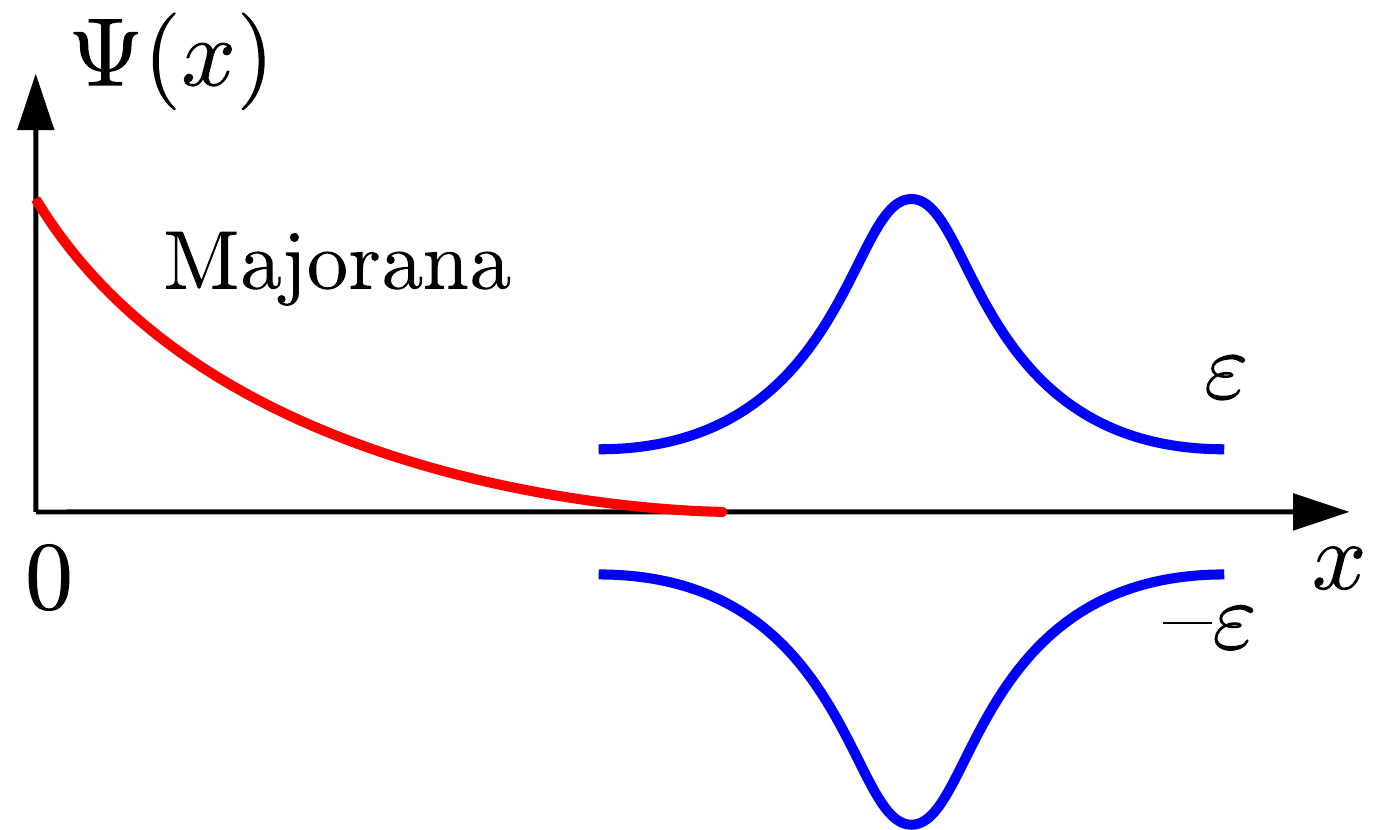}
\caption{(Color online)
A schematic representation of the three localized states whose hybridization
is responsible for $\langle \rho_E(x) \rangle$ at small, but finite energies $E$.}
\label{fig:hybridization}
\end{figure}

We can also remark a similarity of the LDOS profile (\ref{LDOS-infinite})
to the two-point LDOS correlation function in the quasi-one-dimensional wire
\cite{ivanov:09,ivanov:12}. Namely, at low energies, it consists of
two parts: the ``short-range'' part localized at the length scale $\xi$ and
proportional to the single-particle statistics [$\Phi_M(x)$ in our case]
and the ``long-range'' part at the length scale $L_M$. As in the case
of a quasi-one-dimensional wire, this structure may be easily understood
in terms of a Mott hybridization of localized states. The difference from
the two-point-correlation case considered in Ref.~\onlinecite{ivanov:12} is that
now three states hybridize: the Majorana state and the pair of
particle-hole (Bogoliubov--de-Gennes) symmetric \cite{zirnbauer:96,altland:97}
states at energies $\varepsilon$ and $-\varepsilon$
(see Fig.~\ref{fig:hybridization}). The hybridization Hamiltonian involving
these three states has the form
\begin{equation}
H_\textrm{hybrid}=
  \begin{pmatrix}
   \varepsilon & J & 0   \\
   J^*         & 0 & J^* \\
   0           & J & -\varepsilon
  \end{pmatrix}\, ,
\end{equation}
where the {\em typical}\ value of the hybridization
matrix element $J$ is $\Delta_\xi e^{-x/(2\xi)}$
(note that there is no direct hybridization between
the state at energy $\varepsilon$ and its image
in the symmetry class B \cite{altland:97}).
Following the argument of Ref.~\onlinecite{ivanov:12},
we conclude that this hybridization mechanism produces
a crossover between
$\rho_E(x)=0$ and $\rho_E(x)=1$ at $x=L_M$, in
agreement with the result (\ref{LDOS-infinite})
of our calculation. Following the same
argument, the hybridized states reproduce the Majorana
wave function at $x \lesssim \xi$, and therefore
$\langle \rho_E(x) \rangle$ must be proportional
to the average intensity of the Majorana state $\Phi_M(x)$
in this region of $x$, again in agreement with
Eq.~(\ref{LDOS-infinite}).

It is instructive to compare this situation with that in
superconductor--normal-metal (SN) junctions in the symmetry classes
C and D considered in Ref.~\onlinecite{skvortsov:12}. In class D,
there is no hybridization between any state and its particle-hole
mirror image \cite{altland:97}. Since there is no state at $E=0$, this implies
no repulsion from the Fermi level, and $\langle \rho_E(x) \rangle$
is enhanced at low energies at $x \lesssim \xi$
(with $\langle\rho_{E{\to}0}(0)\rangle=3$).
In class C, the particle-hole mirror images hybridize and repel
each other, and therefore $\langle \rho_E(x) \rangle$ is suppressed at $x<L_M/2$
(as opposed to $x<L_M$ in class B), since hybridization involves
an electron (hole) propagating to the interface with the superconductor
and back.
The comparison of the three classes is presented in Table \ref{table:BCD}.

\begin{table}[tb]
\caption{Level repulsion and Mott hybridization in normal-metal--superconductor
junctions of the symmetry classes B, D, and C in the limit of an infinite
wire $L \to \infty$.}
\label{table:BCD}
\begin{ruledtabular}
\begin{tabular}{ccccc}
Class & Majorana & $\alpha$ & hybrid.~length & $\langle\rho_{E{\to}0}(0)\rangle$ \\
\hline
B & Yes & $2$ & $2\xi\ln(\Delta_\xi/E)$ & 1 \\
D & No & $0$ & --- & 3 \\
C & No & $2$ & $\xi\ln(\Delta_\xi/E)$ & 0 \\
\end{tabular}
\end{ruledtabular}
\end{table}

We remark that finite limits $\langle \rho_{E{\to}0}(x) \rangle$
in classes B and D reported in Table \ref{table:BCD} are only achieved if the limit $L\to\infty$
is taken before $E\to0$. If one considers the opposite order of limits, (i.e., $E\to0$
at a finite $L$), then the repulsion from the Fermi level leads to
$\langle \rho_{E{\to}0}(x) \rangle = 0$. In fact, in a finite wire, the hybridization is
only possible at distances $x<L$, which sets the energy scale $E_g \sim \Delta_\xi e^{-L/(2\xi)}$
in class B (whereas $E_g \sim \Delta_\xi e^{-L/\xi}$ in class C) as the typical minimum energy of
hybridized states. Below this energy scale, the density of states is suppressed, which is
reminiscent of a minigap in conventional SN junctions \cite{golubov:89}.
In a finite wire, at very small energies ($E\ll E_g$), the density of states
scales as $\langle\rho_{E{\to}0}(x)\rangle\propto|E|^\alpha$. A finite value of $\alpha$ indicates
repulsion either from the Majorana state or from the Bogoliubov--de-Gennes mirror image: it
equals the number of independent degrees of freedom in the possible hybridization matrix
element between the two repelling states.
The values of $\alpha$ are included in Table \ref{table:BCD} for completeness.
They are also known from the random matrix theory for all symmetry
classes (Refs.~\onlinecite{mehta:91,altland:97,ivanov:02} and references therein).

\paragraph{6. Discussion.---}
To summarize, we have analytically solved the problem of Anderson localization
of a Majorana fermion in a wire with
time-reversal and spin-rotational symmetries broken.
We find that the Majorana level
gets localized in a way qualitatively similar to other electronic states and contributes
to the repulsion of other electron levels from the Fermi energy near the SN interface.
We expect that these qualitative properties are universal: they do not depend on the details of
geometry and on the symmetry class of the wire. Moreover, since the statistics
of the envelope of a single localized wave function in quasi-one-dimensional geometry
is independent of the symmetry class \cite{mirlin:97:00} and since the behavior at the Mott length
scale is determined by the hybridization physics \cite{mott:68:70,ivanov:12},
we believe that our results
(\ref{LDOS-infinite}) and (\ref{Majorana-infinite}) are universally valid also
for wires with orthogonal and symplectic symmetries, as long
as a Majorana level is present at the SN interface.

Our results may be useful for possible qubit designs involving
Majorana fermions localized by disorder. The intensity $\Phi_M(x)$ of the
Majorana state may be probed by tunneling experiments (while the
maximum of the tunneling peak at zero temperature must equal $2e^2/h$,
according to Ref.~\onlinecite{law:09}, its width and temperature
smearing are sensitive to the amplitude  $\Phi_M(x)$ \cite{ioselevich:12}).
Also, the profile of $\Phi_M(x)$ would be relevant for hybridization of Majorana
fermions, if two of them are placed at the opposite ends of the
wire (similarly to the case of a superconducting wire considered
in Ref.~\onlinecite{brouwer:11}).

This work was partially supported by the program ``Quantum mesoscopic
and disordered structures'' of the RAS, RFBR grant No.\ 13-02-01389,
and the German Ministry of Education and Research (BMBF).

\clearpage

\renewcommand{\theequation}{S\arabic{equation}}
\renewcommand{\thepage}{S\arabic{page}}
\setcounter{page}{1}
\setcounter{equation}{0}

\onecolumngrid

\centerline{\large\textbf{Supplemental Material}}
\vspace*{12mm}

\twocolumngrid

Here we provide the details of the calculation leading to the
results (\ref{LDOS-infinite}) and (\ref{Majorana-infinite}) for the LDOS
in the infinite-wire limit. The calculation extensively uses the techniques
developed in Ref.~\onlinecite{ivanov:09}. In the fermionic sector, the calculation
is done perturbatively, while in the bosonic sector we use the
Poisson summation formula involving the scattering matrix
for the Hamiltonian $\tilde H_B$ [Eq.~(\ref{tilde-HB})].
It will be convenient to perform the calculation in the variables $p$ and $q$
introduced below Eq.~(\ref{zero-mode}). We also introduce the dimensionless
coordinate $t=x/\xi$. The main idea of the calculation is to keep
track of the relevant orders in $\kappa$.

The starting point of the calculation are Eqs.\ (\ref{rho-integral}), (\ref{psi-infinite}),
and (\ref{zero-mode}). The Hamiltonians (\ref{tilde-HB}) and (\ref{tilde-HF}) in the variables
$p$ and $q$ read
\begin{eqnarray}
\tilde H_B
& = & \frac{1}{8} \left[ \frac{1}{p} \frac{\partial}{\partial p} p (\kappa^2 - p^2) \frac{\partial}{\partial p}
+ p^2 - \frac{\kappa^2}{2} \right]\, , \\
\tilde H_F
& = & \frac{1}{8} \left[ \frac{1}{q} \frac{\partial}{\partial q} q (q^2 - \kappa^2) \frac{\partial}{\partial q}
- q^2 + \frac{\kappa^2}{2}\right]\, .
\end{eqnarray}

In the fermionic sector, we expand the wave function in powers of $\kappa$ and $q$:
\begin{equation}
 \Psi(q, p; t)
  = \psi_0(p, t) + q^2 \psi_1(p, t) + \kappa^2 \psi_2(p, t) + \ldots\, ,
\end{equation}
which, upon substitution into Eq.~(\ref{rho-integral}), results in
\begin{multline}
 \langle \rho_E(x) \rangle
  =  1 + \frac{1}{2} \Re \int d\lambda_B\, \Big[
        2 \psi_0(p,t) \\
        + \kappa^2 \psi_1(p,t) + 2 \kappa^2 \psi_2(p,t) + \ldots  \Big]\, .
 \label{rho_exp}
\end{multline}
The Hamiltonian $\tilde H_F$ acts as
\begin{multline}
 \tilde H_F \Big[
   \psi_0 + q^2 \psi_1 + \kappa^2 \psi_2
 \Big] \\
  = (q^2 - \kappa^2/2) \Big[
      -\psi_0/8 + \psi_1
    \Big] + o(q^2, \kappa^2)\, ,
\end{multline}
which allows us to represent it by a finite matrix in the basis $(1,q^2,\kappa^2)$:
\begin{equation}
 \tilde H_F \begin{pmatrix}
   \psi_0 \\ \psi_1 \\ \psi_2
 \end{pmatrix}
  = \begin{pmatrix}
      0 & 0 & 0 \\
      -1/8 & 1 & 0 \\
      1/16 & -1/2 & 0
    \end{pmatrix} \begin{pmatrix}
   \psi_0 \\ \psi_1 \\ \psi_2
    \end{pmatrix} + o(q^2, \kappa^2)\, .
\end{equation}
Exponentiating this matrix, we obtain
\begin{multline}
 e^{-2 \tilde H_F t} \Big[
   \psi_0 + q^2 \psi_1 + \kappa^2 \psi_2
 \Big] \\
  = \psi_0
    +q^2 \left[ \frac{1 - e^{-2t}}{8} \psi_0 + e^{-2t} \psi_1 \right] \\
    +\kappa^2 \left[ \frac{e^{-2t} - 1}{16} \psi_0 + \frac{1 - e^{-2t}}{2} \psi_1 + \psi_2 \right]
    + o(q^2, \kappa^2)\, .
\end{multline}

Applying this operation to the zero-mode wave function
\begin{equation}
 \Psi_0(q,p)
  = p K_1(p) + q^2 \left( \frac{K_0(p)}{2} + \frac{p K_1(p)}{4} \right)
  + o(q^2) 
\end{equation}
yields
\begin{multline}
 e^{-2 \tilde H_F t} \Psi_0(q,p)
  = p K_1(p) \\
    +q^2 \left[ \frac{1 + e^{-2t}}{8} p K_1(p) + \frac{e^{-2t}}{2} K_0(p) \right] \\
    +\kappa^2 \left[ \frac{1 - e^{-2t}}{16} p K_1(p) + \frac{1 - e^{-2t}}{4} K_0(p) \right]
    + o(q^2, \kappa^2)\, .
\end{multline}
Inserting this result into Eq.\ (\ref{rho_exp}), we obtain
\begin{multline}
 \langle \rho_E (x) \rangle
  =  1 + \Re \int d\lambda_B\, e^{-2 \tilde H_B t} \Big(
        p K_1(p) \\
      + \frac{\kappa^2}{8} \big[ p K_1(p) + 2 K_0(p) \big]
      \Big)\, .
 \label{rho_B}
\end{multline}
This equation is valid up to corrections of order $o(\kappa^2)$ in the
integrand, which results in corrections of order $o(1)$ for
$\langle \rho_E (x) \rangle$ (the integration over $\lambda_B$ extends to
$\lambda_B \sim \kappa^{-2}$ and thus brings in a large $\kappa^{-2}$ factor).
It remains now to calculate the evolution with respect to the bosonic part of the
Hamiltonian $\tilde H_B$.

We further calculate the following two matrix elements
\begin{equation}
  \begin{Bmatrix} M_1 \\ M_0 \end{Bmatrix}
  = 
  \int d\lambda_B\, e^{-2 \tilde H_B t} \begin{Bmatrix} p K_1(p) \\ K_0(p) \end{Bmatrix}
\end{equation}
using the method developed in Ref.\ \onlinecite{ivanov:09}.
From Eq.~(\ref{rho_B}) we see that the integral involving $p K_1(p)$ should be calculated up to
two leading terms in small energy expansion, $O(1/\kappa^2)$ and $O(1)$, while for the second integral it suffices to extract only the main
$O(1/\kappa^2)$ asymptotics.

We use the general expansion of the evolution operator in the eigenfunctions of $\tilde H_B$
\begin{multline}
  \begin{Bmatrix} M_1 \\ M_0 \end{Bmatrix}
  = \sum_k \frac{\langle 1 | \phi_k \rangle}{\langle \phi_k | \phi_k \rangle} e^{-2 E_k t}
    \begin{Bmatrix} \langle \phi_k | p K_1(p) \rangle \\ \langle \phi_k | K_0(p) \rangle \end{Bmatrix}
\end{multline}
and pick all the necessary ingredients from Ref.\ \onlinecite{ivanov:09}:
\begin{align}
&  \langle \phi_k | p K_1(p) \rangle
  = \frac{\pi}{\kappa \cosh \pi k} \left[
      1 + 4 k^2 - \frac{\kappa^2}{2} + O(\kappa^4)
    \right] , \\
& \langle \phi_k | K_0(p) \rangle
  = \frac{2 \pi}{\kappa \cosh \pi k} \left[
      1 + O(\kappa^4)
    \right] , \\
& \langle \phi_k | \phi_k \rangle
  = -\frac{i}{2\pi} \coth \pi k \frac{\partial \ln S(k)}{\partial k} \left[
      \frac{1}{k} + O(\kappa^4)
    \right] , \\
&  \sum_k \cdots
  = -\frac{i}{2\pi} \sum_{n = -\infty} ^ {+\infty} \int_0^\infty dk\; S^n(k) \frac{\partial \ln S(k)}{\partial k}
  \cdots , \\
&  S(k)
  = \left( \frac{\kappa^2}{4} \right)^{-2ik} \left[
      \frac{\Gamma(2ik) \Gamma(1/2 - ik)}{\Gamma(-2ik) \Gamma(1/2 + ik)}
    \right]^2 \nonumber \\
& \hspace{5cm}    \times \left[
      1 + O(\kappa^4)
    \right] , \\
& E_k
  = \frac{1}{8} + \frac{k^2}{2} + O(\kappa^4) .
\end{align}
The only missing piece is $\langle 1 | \phi_k \rangle$.
It can be easily computed along the lines of Appendix C of Ref.\ \onlinecite{ivanov:09}:
\begin{equation}
 \langle 1 | \phi_k \rangle
  = \int d\lambda_B\, \phi_k
  = \frac{4}{\kappa} \left[
      1 + O(\kappa^4)
    \right].
\end{equation}

Putting everything together, we obtain
\begin{multline}
  M_1
  = 4\pi \int_0^\infty k\, dk\, \frac{\sinh \pi k}{\cosh^2 \pi k} \sum_{n = -\infty} ^ {+\infty} S^n(k) \\
  \times \left[
      \frac{1 + 4 k^2}{\kappa^2} - \frac{1}{2} + O(\kappa^2)
    \right] e^{-t(k^2 + 1/4)}.
\end{multline}
The term with $n = 0$ rapidly converges as an integral over real $k$, while $n = \pm 1$
terms oscillate and are determined by the
competition between the contributions of the pole at $k = i/2$ and of the saddle point $k_* = i t_M/2t$
with $t_M = 4 \ln(2/\kappa)$. The pole contribution is
dominant at $t \ll t_M$ and yields $-1$. The saddle point $k_*$ becomes
important at $t \sim t_M$ and provides a step-like ``erf term''. The result is
\begin{multline}
M_1
  = -\frac{1}{2} + \frac{1}{2} \erf \left( \frac{t - t_M}{2 \sqrt{t}} \right) \\
    + 4\pi \int_0^\infty k\, dk\, \frac{\sinh \pi k}{\cosh^2 \pi k} \left[
      \frac{1 + 4 k^2}{\kappa^2} - \frac{1}{2}
    \right] e^{-t(k^2 + 1/4)} \\ + O(\kappa^2).
 \label{int_pK1}
\end{multline}

Evolution of $K_0(p)$ is easier since we need only the leading term.
Setting $n = 0$, we find
\begin{multline}
  M_0
  = \frac{8\pi}{\kappa^2} \int_0^\infty k\; dk\; \frac{\sinh \pi k}{\cosh^2 \pi k} e^{-t(k^2 + 1/4)} + O(1).
 \label{int_K0}
\end{multline}

The results  (\ref{LDOS-infinite}) and (\ref{Majorana-infinite}) for the LDOS are now obtained  in
a straightforward way by substituting the integrals
(\ref{int_pK1}) and (\ref{int_K0}) into Eq.~(\ref{rho_B}). The
Majorana delta peak arises from $\Re (1/\kappa^2) = (\pi/8) \delta(E/\Delta_\xi)$.


\begin{thebibliography}{99}
\bibitem{com-delta}
The $1/2$ factor in the definition of the level spacing $\delta$ takes into account the doubling
of levels due to Andreev reflections and is included for consistency of notations
with Refs.~\cite{altland:97,ivanov:02,skvortsov:12}.
\end{thebibliography}
\end{document}